\documentclass[conference]{IEEEtran}
\usepackage{blindtext, graphicx}
\ifCLASSINFOpdf
\else
\fi

\usepackage{amsmath}
\usepackage[noend]{algpseudocode}
\usepackage{algorithm}
\usepackage{amssymb}

\begin{document}
%
\title{Evolving Social Networks via Friend Recommendations}

\author{\IEEEauthorblockN{Amit Kumar Verma}
\IEEEauthorblockA{Department of Computer Science and Engineering\\
National Institute of Technology Meghalaya, Shillong \\
India - 793003\\
Email: mt4descentis@gmail.com}
\and
\IEEEauthorblockN{Manjish Pal}
\IEEEauthorblockA{Department of Computer Science and Engineering\\
National Institute of Technology Meghalaya, Shillong \\
India - 793003\\
Email: manjishster@gmail.com }
}


%


\maketitle

\begin{abstract}
A social network grows over a period of time with the formation of new connections and relations. In recent years we have witnessed a massive growth of online social networks like Facebook, Twitter etc. So it has become a problem of extreme importance to know the destiny of these networks. Thus predicting the evolution of a social network is a question of extreme importance. A good model for evolution of a social network can help in understanding the properties responsible for the changes occurring in a network structure. In this paper we propose such a model for evolution of social networks. We model the social network as an undirected graph where nodes represent people and edges represent the \emph{friendship} between them. We define the evolution process as a set of rules which resembles very closely to how a social network grows in real life. We simulate the evolution process and show, how starting from an initial network, a network evolves using this model. We also discuss how our model can be used to model various complex social networks other than online social networks like political networks, various organizations etc..       
\end{abstract}


%
\IEEEpeerreviewmaketitle

\section{Introduction}
The analysis of online Social Networks has allowed us to answer many questions regarding the characteristics of network and how the network changes. Modeling the evolution of a social network and predicting the structure of the future network is a complex problem. Social networks grow and change quickly over time with the addition of new edges, signifying the appearance of new interactions/relations in the underlying social structure. In this work,  we consider a social network as a network comprising of members which are connected in the network by the ``friendship'' relation. We try to understand the mechanisms by which the social network evolves over time and using this information we design a model which allows us to predict the structure of the future network.\\

In order to model the evolution of a Social Network we have to know the network characteristics, as in many evolution studies, the underlying process for network change is assumed to be centered at the behavioral characteristics of the network members \cite{J.Coleman}. In a social network, network members  tend to 'choose' their friends by comparing relevant individual characteristics of the others with their own. A fundamental finding in many choice networks is that social members with similar characteristics are more often connected with one another than with more dissimilar ones. This is known as the "similarity effect" in social networks \cite{schachter}. So we can infer that for any relationship between the network members of a social network there must be some common characteristics between those members which lead to that relation. This basic idea can be used to design the evolution model. In our work we call these common characteristics as \emph{factors}. For example in online social networks like Facebook we can observe certain \emph{factors} namely frequency with which friends tag each other in their posts, place where they live, work place (school, office, university etc.), common interests (movies, songs, books) etc. These \emph{factors} signify how similar the members of the social network are. Considering more \emph{factors} we can come up with a model that can provide a more precise measure of the similarity amongst the individuals. It has also been shown in \cite{Leenders} that gender plays an important role in deciding the level of friendship in people thus gender can also be considered as a \emph{factor}. The strength of the friendship depends on the number of common \emph{factors} as well as the weight of these common \emph{factors}.  It means that we can associate the term "quality" with these relationships which shows how strong is the relationship among the members. \\
In the following section we will discuss some prior work regarding evolution models. Subsequently, we will describe the model of evolution proposed by us.

\subsection{Related Work}
Several researchers have turned their attention to the evolution of social networks at a global scale. For example networks become denser over time, in the sense that the number of edges grows super-linearly with number of nodes \cite{Leskovec1}. In this paper they reported that the network diameter often shrinks over time, in contrast to the conventional concept that such distance measures should increase slowly as a function of the number of nodes. Some efforts has also been made in the direction to find the properties responsible for the network evolution. A variety of network formation strategies were investigated showing that \emph{edge locality} plays a critical role in network evolution \cite{Leskovec2}. Many models have been designed to predict the links in social networks for example in \cite{Berlingerio} they introduced the notion of graph evolution rules in which they developed Graph Evolution Rule Miner(GERM) software to extract the rules responsible for network evolution and applied these rules to predict the future network. In the direction of basic principles responsible for social network evolution researchers have shown that the most important characteristic of social evolution is that the outcome of evolution process is not the result of central authority but are the consequences of the simultaneous choices of persons \cite{Stokman}\cite{Zeggelink1}\cite{Zeggelink2}. Social actors try to realize their own goals by choosing between behavioral alternatives that are available to them under certain restriction \cite{Wippler}. In \cite{riitta} they have reviewed, classified and compared different models of social networks. They have classified these models into two main categories, first in which the addition of new links depends on the local network structure(\emph{Network Evolution Models}), and second in which probability of each link existing depends only on nodal attributes (\emph{Nodal Attribute Models}). In their work, they have shown that \emph{Nodal Attribute Models} produce very clear community structure. In next section we will discuss about our work and will describe our evolution model. 

\section{Our Contribution}
In this section we discuss about the evolution model that we have proposed. The basic idea that we have used to design this evolution model is the \emph{transitive} property among the relationships i.e if a person \emph{A} is connected to a person \emph{B} who in turn is connected to a person \emph{C}, then there is high possibility that person \emph{A} and \emph{C} will be connected. We have designed the evolution model which uses this transitive property to evolve the network.  In our model we represent the network as undirected graph such that nodes represent people and edges represent connections between them. These connections can be of any form of friendship relation like "friends", "acquaintance", "co-workers" etc. 
\subsection{Preliminaries}
We have modeled the social network as a graph $G = (V,E)$ where the nodes($V$) represent the people present in the network and an edge $e = (u,v) \in E$ represents the friendship relation between them. As we have already discussed that for any social relation there must be some common characteristics (\emph{factors}) which decides these relations. For this purpose we have associated \emph{factors} with each edge. For simplicity, in our model we have assumed that the number of \emph{factors} is finite. With each edge we associate certain \emph{factors} which represent the common attributes between the nodes which are connected by that edge.  Let $F$ be a finite set of \emph{factors} s.t $F \subset \mathbb{N}$ and $F_j \subseteq F$ be the \emph{factors} of an edge $e_j$ for $j = 1, 2, 3, ..., m$ where $m$ is the total number of edges. To decide the "quality" of \emph{friendship} we have associated \emph{score} values with these \emph{factors}. These values will decide at what degree the \emph{factors} between the people are similar and the cumulative \emph{score} value will be the measure of "quality" of \emph{friendship}. For each factor in set $F_j$ a \emph{score} value $s$ is given to it. For example if ${f_1, f_2, f_3, f_4, ...., f_k}$ is a set of \emph{factors} for an edge $e_j$ then ${s_1, s_2, s_3, s_4, ...., s_k}$ are the \emph{score} values associated with these \emph{factors}. The cumulative \emph{score} value for each set of \emph{factors} $F_j$ on edge $e_j$ is denoted by $S_j$ which is equal to the $\sum_{i=1}^k s_i$. Now we have defined the structure of our graph $G = (V,E)$ in which $V$ is the set of nodes and $E$ is the set of edges and each edge $e_j$ is associated with a set of \emph{factors} $F_j$ and a cumulative \emph{score} value $S_j$ for $j$ = $0,1,2, 3, ...., m$ where $m$ is total number of edges in graph $G$. In next section we will describe the evolution process for our model.

\subsection{Process of Evolution}
Let at time $i =0$, initial graph be $G_0$ and set of score values be $score_0$ which is the union of the all the scores of the edges present in $G_0$. At every time step we will try to add edges between the nodes according to the evolution rules. Suppose at time stamp $i$, $G_i = (V, E_i)$ is the graph evolved from the initial graph then for all $i \geq 0$, $G_i = (V,E_i)$ is subgraph of $G_{i+1} = (V, E_{i+1})$ i.e. $E_{i} \subseteq E_{i+1}$. According to our rules of evolution $E_{i+1}$ is obtained from $E_i$. The set of scores of edges of $G_i$ is denoted by $S_i = \{s_{uv} | (u,v) \in E_i \}$. In this model, once an edge is added to the graph during the evolution process, it is never deleted in any future time-stamp. Also the number of nodes in the graph is fixed. We assume that the set of factors $F_e$ for an edge $e$ will never change during the evolution process, i.e  if $F_{e}$ is the set of \emph{factors} for an edge $e$ at time step $i$ then at any time step $k > i$, in $G_k$, $F_e$ will remain to be the set of \emph{factors} for the edge $e$. We have also considered a model in which we are allowed to add some nodes in the graph, but once added it will never be removed. In our next section we will describe the evolution rules on the basis of which the evolution of the network is done.

\begin{figure}[htb]
\centering
\includegraphics[scale=0.55]{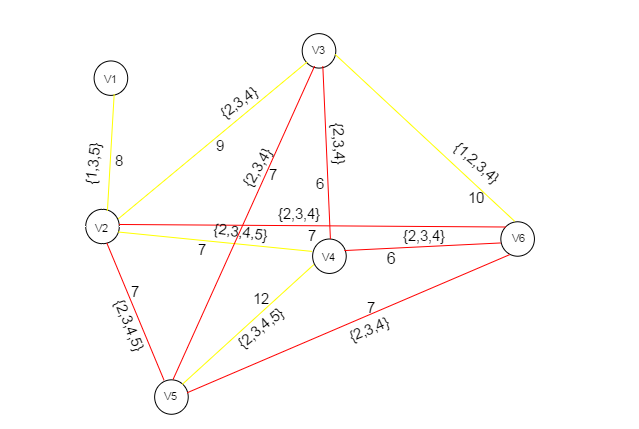} 
\caption{An example with six nodes}
\label{fig:label} 
\end{figure}

\subsection{Rules of Evolution}
In this section we present the rules of evolution of $(G_i,score_i)$ which lead
to some interesting properties of the social network graph. Starting from
an initial social network these rules help in predicting the state of
the social network after several steps. Our main assumption in the evolution 
process is that to add a new edge $(u,v)$ say at the $i^{th}$ step 
there has to be another vertex $w$ such that $(u,w)$ and $(w,v)$ are already 
existing edges in the graph.  This modelling is based on a general
real life observation that in order for two strangers to become \emph{friends} there is usually a
common friend. We also ensure that this edge is added if the
\emph{cumulative score} is more than a particular threshold $t$. We define this cumulative score 
by first taking the intersection of the sets $F_{(u,w)}$ and $F_{(w,v)}$ and then computing
the maximum of $\frac{|F_{(u,w)} \bigcap F_{(w,v)}|}{2} \cdot \left( \frac{S_{(u,w)}}{|F_{(u,w)}|} + \frac{S_{(w,v)}}{|F_{(w,v)}|} \right)$
over all possible $w \in V$. Notice that in order to define the cumulative score we
have considered the \emph{arithmetic mean} of the values  $\frac{S_{(u,w)}}{|F_{(u,w)}|}$ and $\frac{S_{(w,v)}}{|F_{(w,v)}|}$
and scaled them by the cardinality of the set of common \emph{factors}. The intuitive reason for this score function is as follows: 
\begin{itemize}
\item The quantity  $\frac{S_{(u,w)}}{|F_{(u,w)}|} $ represents the score of a \emph{factor} for a particular
edge and thus by taking an arithmetic mean of the two quantities we get the average score of 
a \emph{factor} that is common in both $F_{(u,w)}$ and $F_{(w,v)}$.

\item Multiplying the arithmetic mean by $|F_{(u,w)} \bigcap F_{(w,v)}|$ scales this weight by 
the number of \emph{factors} common to both $F_{(u,w)}$ and $F_{(w,v)}$.
\end{itemize}

Notice that instead of taking arithmetic mean of the two quantities we can also
take the geometric mean or harmonic mean to get an average score.
Also despite the fact we can recommend an edge that has a cumulative score of an more than the threshold 
there is still a chance that two people might decide to not be \emph{friends} because of
some random uncorrelated event. Thus in order to take care of this issue we make our evolution rules randomized i.e. 
even though an edge has a cumulative score more than the threshold it is added
in the evolution process with probability $p$ and rejected with probability $1-p$.
In the following we formally describe the rules of evolution of the network.

\begin{algorithm}
\caption{Evolution Process}
\begin{algorithmic}
\State $i \gets 0$
\While{no more edges can be added }
\State for each $(u,v)$ is not added, and $\forall$ $w$ such that $(u,w), (w,v) \in E_i$ find,
\State \[z = \arg\max_w(\frac{k}{2} \cdot (\frac{s_{uw}}{k_{uw}} + \frac{s_{wv}}{k_{wv}}))   \mbox{ where } k = F_{uw} \bigcap F_{wv}. \]
\If {$(\frac{k}{2} \cdot (\frac{s_{uz}}{k_{uz}} + \frac{s_{zv}}{k_{zv}}))_w > t$}
\State add $(u,v)$ with probability $p$ and discard with probability $1-p$.
\EndIf
\State $i \gets i+1$
\EndWhile
\end{algorithmic}
\end{algorithm}

In figure \ref{fig:label} we can see the example of evolution. In this figure the network is made up of six nodes with starting connections marked as yellow, and after evolving this network we get the connections which are marked in red. For example $V_2$ is connected to $V_3$ with the set of \emph{factors} $F_{V_2,V_3}$ as \{2,3,4\} and \emph{score} value $s_{V_2,V_3}$ as 9 and $V_3$ is connected to $V_6$ with the set of \emph{factors} $F_{V_3,V_6}$ as \{1,2,3,4\} and \emph{score} value $s_{V_3,V_3}$ as 10. So after applying the score function we get the intersection of \emph{factors} as \{2,3,4\} and score value as 7. Since this calculated score value is greater than the threshold value (which is 6), this new link $V_2$ to $V_6$ is added as a new connection. In our evolution model we add the edge with some probability. This process will repeat until there are no new edges that can be added.      

\section{Experiments and Result}
To implement our evolution model we have considered some assumptions. The input to the evolution process is a graph which we called as "initial graph", so to produce these initial graphs we designed a program which produces random graph by putting random edges between the nodes. So each time we require a graph we run this program to get an initial graph which will act as input to our evolution model. As we have described that the set of \emph{factors} which are responsible for addition of edges can vary from person to person, we have assigned \emph{factors} on each edge randomly. So on each edge $e_j$ we have assigned a set of \emph{factors} $F_j$ randomly from the set of total \emph{factors} $F$. To decide the quality of "Friendship" we have assigned score values on each edge randomly as well.

\begin{figure}[htb]
\centering
\includegraphics[scale=0.35]{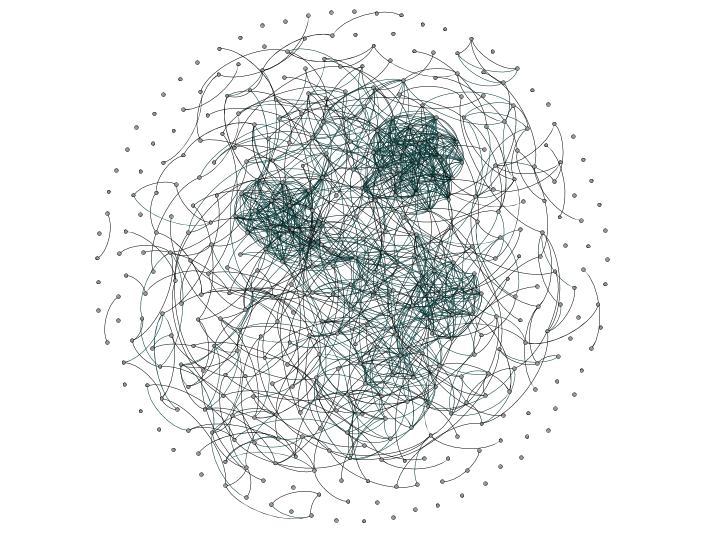} 
\caption{Evolution process on 400 nodes using arithmetic mean}
\label{fig:first} 
\end{figure}

Consider an experiment in which the initial graph has 400 nodes and randomly placed edges. On each edge $e_j$ the set of \emph{factors} $F_j$ that has been assigned is taken randomly from the set of all \emph{factors} $F$ = \{$1, 2, 3, ..., 8$\}. The score value on each edge $e_j$ has been also taken randomly from the set of all score values $S$ = \{$1, 2, 3, 4, ..., 16$\}.  On this initial graph we apply our evolution process to get the evolved graph. To visualize the graph we have used Gephi tool.  In figure \ref{fig:first} we can see the evolved graph.

\subsection{Experiment with different mean }

\begin{figure}[htb]
\centering
\includegraphics[scale=0.35]{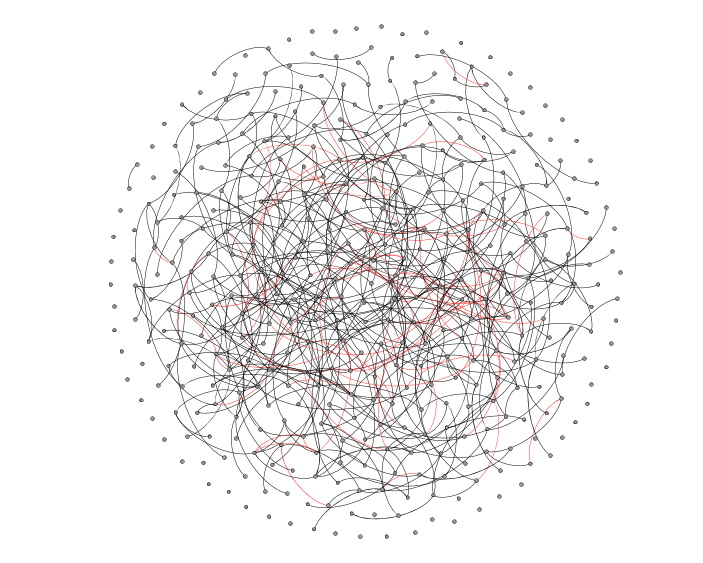} 
\caption{Evolution process on 400 nodes using geometric mean}
\label{fig:second} 
\end{figure}

In the previous experiment we have calculated the cumulative \emph{score} value by taking the \emph{arithmetic mean} of the individual \emph{score} values. The new connections are made by calculating the cumulative \emph{score} values of the common \emph{factors} of a transitive relation. We consider another experiment on 400 nodes again with a random initial graph, but instead of taking the arithmetic mean of the score values we take their geometric mean to compute the cumulative \emph{score} values.  In figure \ref{fig:second} the evolution does not show the formation of communities as was established by the previous experiment.
 
 \subsection{Experiments with different number of nodes}
 In this section we will show the experiment results for different number of nodes.
 
\begin{figure}[htb]
\centering
\includegraphics[scale=0.35]{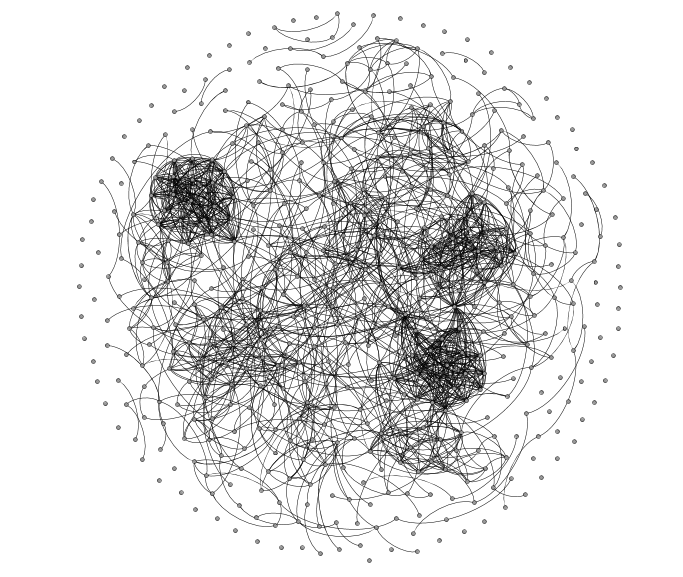} 
\caption{Evolution process on 500 nodes}
\label{fig:third} 
\end{figure}

\begin{figure}[htb]
\centering
\includegraphics[scale=0.35]{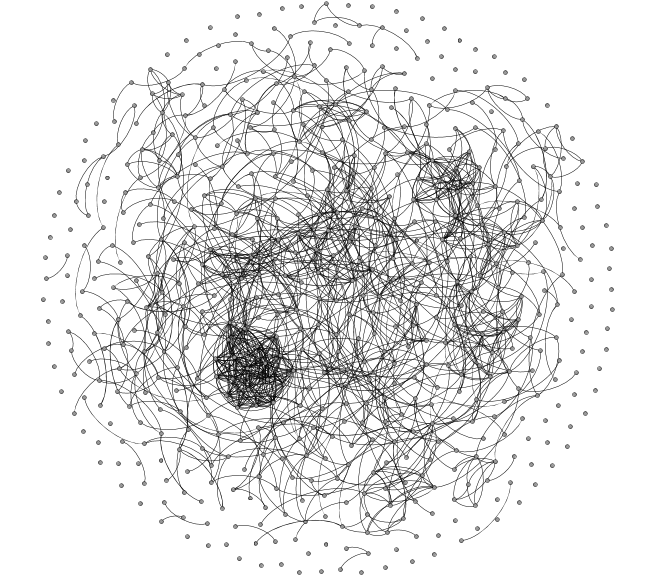} 
\caption{Evolution process on 600 nodes}
\label{fig:forth} 
\end{figure}

\begin{figure}[ht]
\centering
\includegraphics[scale=0.35]{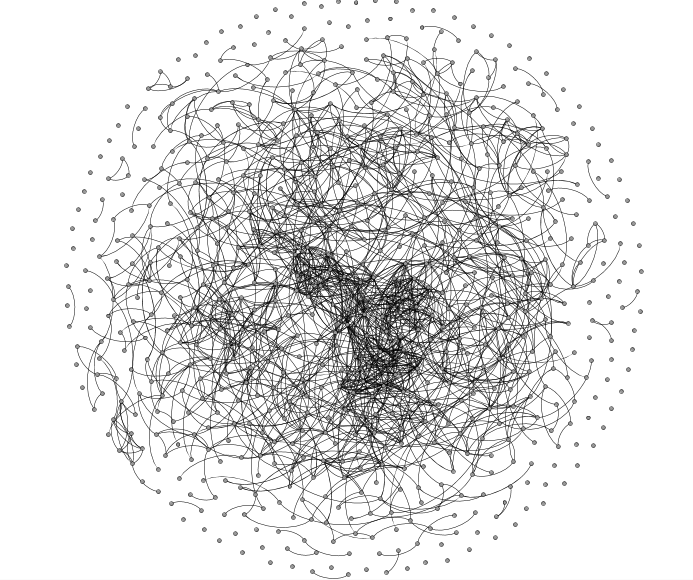} 
\caption{Evolution process on 700 nodes}
\label{fig:fifth} 
\end{figure}

\subsection{An iterative evolution process}
We have proposed another extension of our model in which rather than applying the evolution process to fixed number of nodes we are iteratively adding more nodes to the graph and applying the evolution process. This process will not terminate as nodes will be added at every stage of evolution so we have taken the snapshot of graph at time $T$. In this process random number of nodes from a new set of nodes $V_{new}$ are added at every time step $T$. These nodes are randomly connected to some nodes which are already present in the network and on this modified graph the evolution process is applied. So at some time step $T$ we produce the graph which can be considered as a snapshot of the evolution process. The following algorithm shows the iterative evolution process.

\begin{algorithm}[t]
\caption{Iterative Evolution Process}
\begin{algorithmic}
\While{no more new nodes are added}
\State $i \gets 0$
\While{no more edges can be added to $G_i$}
\State for each $(u,v)$ is not added, and $\forall$ $w$ such that $(u,w), (w,v) \in E_i$ find,
\State \[z = \arg\max_w(\frac{k}{2} \cdot (\frac{s_{uw}}{k_{uw}} + \frac{s_{wv}}{k_{wv}}))   \mbox{ where } k = F_{uw} \bigcap F_{wv}. \]
\If {$(\frac{k}{2} \cdot (\frac{s_{uz}}{k_{uz}} + \frac{s_{zv}}{k_{zv}}))_w > t$}
\State add $(u,v)$ with probability $p$ and discard with probability $1-p$.
\EndIf
\State $i \gets i+1$
\EndWhile
\State add random set of nodes $V_{new}$ and a random set of edges connecting to $G_i$.
\EndWhile
\end{algorithmic}
\end{algorithm}

\begin{figure}[htb]
\centering
\includegraphics[scale=0.4]{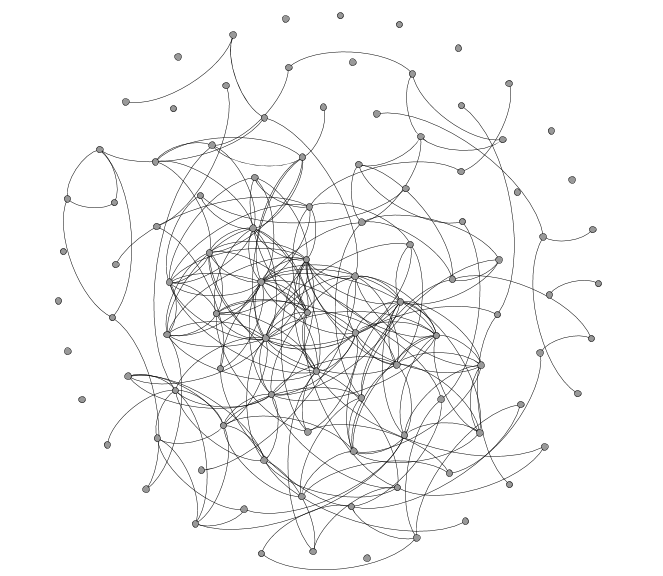} 
\caption{Iterative Evolution process on 100 nodes}
\label{fig:it} 
\end{figure}

\begin{figure}[htb]
\centering
\includegraphics[scale=0.4]{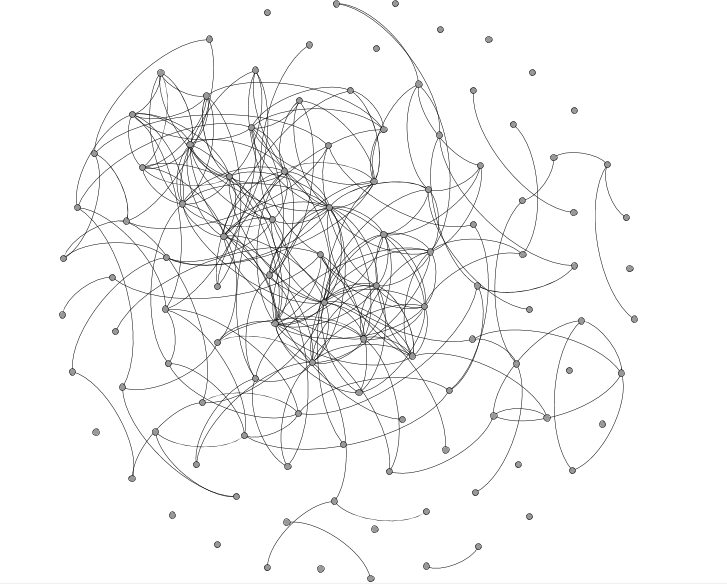} 
\caption{Evolved graph at time step $T+k$}
\label{fig:it_2} 
\end{figure}

We have implemented the iterative evolution process on 100 nodes. We can see in figure \ref{fig:it} the graph on which the iterative evolution process has been applied. In this experiment we have taken the set of nodes $V_{new}$ as 20 from which random number of nodes are taken and are added to the graph at every time step, and the evolution process continues with the modified graph. Figure \ref{fig:it} shows the initial graph to which some random nodes are connected to it. Figure \ref{fig:it_2} is the graph obtained after evolution when 10 nodes get added random to the graph in \ref{fig:it} at some time step.  We can also observe formation of communities in iterative evolution process as shown by the networks evolved through the evolution process with fixed number of nodes.   

\section{Conclusion}
In this work we propose a model of evolution of social networks
based on the transitive property of the growth of friendship relation.
Our model is based on predicting the similarity of two people which 
satisfy the transitive property. We characterize a particular relation i.e.
an edge in the social network graph, by a number of sociopsychological \emph{factors} and a
score value which measure the strength of the relationship. 
We observe that if we use a specific formulation to derive the strength
of a recommended friendship using arithmetic mean, dense communities
are formed which is a characteristic of real life social networks. Changing the nature
of the formulation using other means doesn't show formation communities. 
To the best of our knowledge such a model has not been proposed yet.
We suspect that this model can be useful in understanding the evolution of
complex social networks like online social networks, political networks, corporate networks etc.
As part of future work, it would be interesting to see whether our model can
make accurate predictions for real life social networks in terms of formations of communities.
It would also be interesting to prove theoretical guarantees on the results obtained
in this paper which would in turn be an in depth study of the theoretical underpinnings
of this work.

\begin{IEEEbiography}[{\includegraphics[width=1in,height=1.25in,clip,keepaspectratio]{picture}}]{John Doe}
\blindtext
\end{IEEEbiography}




\end{document}